*Research Article*

# Application of Wireless Sensor Networks for Indoor Temperature Regulation

**Biljana Risteska Stojkoska,[1] Andrijana Popovska Avramova,[2] and Periklis Chatzimisios[3]**

[1] *Faculty of Computer Science and Engineering, Saints Cyril and Methodius University, 1000 Skopje, The Former Yugoslav Republic of Macedonia*
[2] *Department of Photonics Engineering, Technical University of Denmark, 2800 Kongens Lyngby, Denmark*
[3] *Department of Informatics, Alexander TEI of Thessaloniki, 57400 Thessaloniki, Greece*

Correspondence should be addressed to Biljana Risteska Stojkoska; biljanastojkoska@yahoo.com

Wireless sensor networks take a major part in our everyday lives by enhancing systems for home automation, healthcare, temperature control, energy consumption monitoring, and so forth. In this paper we focus on a system used for temperature regulation for residential, educational, industrial, and commercial premises, and so forth. We propose a framework for indoor temperature regulation and optimization using wireless sensor networks based on ZigBee platform. This paper considers architectural design of the system, as well as implementation guidelines. The proposed system favors methods that provide energy savings by reducing the amount of data transmissions through the network. Furthermore, the framework explores techniques for localization, such that the location of the nodes can be used by algorithms that regulate temperature settings.

## 1. Introduction

Wireless sensor networks (WSNs) are able to efficiently sense various parameters with high accuracy and low power consumption. The development of sensors and networks based on sensor nodes has impacted and changed our everyday life. Engaging WSNs in home and industrial monitoring systems, medicine and healthcare systems, entertainment, education, and so forth, has enlightened and improved the concept of modern living. A wireless sensor network consists of three major elements [1]: sensor unit (used to take measurements), computing unit (used to process data), and communication unit (used to enable communication among the wireless nodes). Different radio technologies can be used for communication, such as ZigBee, Wi-Fi, Bluetooth, and Global Systems for Mobile Communications (GSM). ZigBee as an emerging technology has been proven to make WSN self-configurable and self-healing while operating at low power consumption [2], a feature that is very important for wireless sensors.

Intelligent smart home frameworks have been proposed recently by the research community in [3, 4]. The proposed systems are used to monitor and report different parameters in a home environment such as temperature, humidity, and light, as well as controlling different electrical devices for lightning, air conditioning, or heating. In [4], energy optimization is based on a dynamic programming algorithm that controls the usage of energy and sells it back to the smart grid. In [5], the authors have proposed a prototype system for temperature monitoring in a university campus. The purpose of this system is to provide optimal management of the cooling system in order to reduce the power consumption. The system consists of the client part with web-based interface and MySQL database and two types of nodes: coordinator node that is responsible for data gathering and terminal nodes that measure temperature, humidity, and light intensity. This prototype application is implemented on a centralized network (utilizing a star topology). The maximum distance between the coordinator and the terminal nodes is 140 meters in open space or 40 meters when obstacles are present. Both terminal and coordinator nodes use the same microcontroller, which is based on the Arduino [6] boards and XBee [7] communication modules that support the ZigBee standard. This system has several drawbacks. Centralized network could not cover campus taking into account the limitations of the maximum distance between the terminal nodes and the coordinator. The software does not support a coordinator to store data in a MySQL database, so the idea that they should be stored at the client's side is an inappropriate solution. Arduino microcontrollers are expensive solution for WSN because of their complexity, which is absolutely unnecessary for nodes that have a primitive task to measure three physical parameters. In [8], a WSN system is proposed to control temperature. Unlike previous approaches that focused on the design of the network, in [8] the emphasis is placed on data processing. The authors propose an analysis of temperature readings using the variogram in order to make a prediction of the temperature at each possible location in the room. In [9], a WSN based on star-topology is used for temperature monitoring in a greenhouse. Since radio range of the nodes is 100 m, nodes can directly send their data to the base station. The authors in [10] present a web-based WSN interface that uses state-of-the-art technologies for efficient habitat monitoring. The system is based on Mica motes that are in interaction with remote users. The authors in [11] have installed a modular and extensible WSN in a test and reference household called VILLASMART. The energetic behavior of the building is modeled using indoor and outdoor WSN readings (air and water temperature, solar radiation sensor, weather conditions and power consumption information). Grey-box estimation method is used for model parameter determination, thus more precise predictions of the indoor temperature are achieved.

Our work proposes a wireless sensor network framework for indoor temperature regulation (WSN-FITR). Homes, classrooms, and halls are often heated up by a number of temperature controlled heaters. Users are usually not interested in controlling the temperature at each separate heater. The radiators, for example, are normally located just above the floor or below windows and at the room's walls. Furthermore, the measurements do not show the real room temperature as the temperature sensors are located just next to the heaters. Neighbouring rooms with own heating elements also influence the temperature in the controlled room. This paper presents detailed guidelines for efficient integration of WSN into intelligent system for indoor temperature regulation. We have discussed different WSN topologies regardless of the deployment area. This paper shows how the node localization methods can be used for room temperature optimization in order to provide the most optimal tradeoff between the time it takes to reach the wanted temperature at a specific part of the room and energy consumption. Even more, reduction of the required data transmission through prediction methods is considered, which is important in order to increase the battery life of the nodes and to extend the network lifetime.

The rest of this paper is organized as follows. In the Section 2, the framework architecture of WSN-FITR is presented, while Section 3 provides an overview of ZigBee and network topology of the proposed framework. Section 4 gives a detailed explanation of techniques for indoor localization and clustering with respect to characteristics of the environment where WSN is deployed. Section 5 compares techniques for data reduction in WSN. Finally, we conclude this paper in the last section.

## 2. System Architecture

The proposed WSN-FITR system for indoor temperature regulation has a simple system architecture (Figure 1). The wireless nodes deployed in each room are programmed to monitor the temperature. There are two types of nodes: sensor regulators and temperature controllers which are interconnected in a ZigBee network. Sensor-regulator nodes take samples at predefined intervals and send data to the controller node. The controller node is responsible for collecting and processing sensor readings from its regulator nodes. It actually serves as a gateway that forwards data to the central (base) station. If the distance between the controller and the base station is long, then the data will be forwarded using multihop communication protocol. This data is stored in a database and can be analyzed and processed on demand. Users can access the data using smart phones or desktops via the Internet.

Sensor regulators perform measurement and report local temperature readings to the controller. They are attached to a device and regulate its action (e.g., increase/decrease heating) in order to reach a certain temperature. Typical devices and their corresponding actions are the following:

(i) heating bodies, such as central heating radiators, electric radiators, or fans. The regulator can increase/decrease the heating by using the valve controller,

(ii) air conditioners, which can be based on a fan. The regulator can increase/decrease the cooling volume in order to reach a certain temperature,

(iii) air flow, such as central air flow ventilation. These devices can regulate the degree of air circulation,

(iv) window shutters, such as outside curtains. By rising/lowering the curtains the influence of the sun energy can be regulated.

The temperature controller nodes are considered more powerful than the sensor regulators. They are expected to have more advanced capabilities: memory, processing unit, and steady energy supply. They should be located higher up in the room and away from all the heaters and windows in a location that better represents the room's temperature in order to measure the most relevant temperature. Additionally they are used to control the temperature at the premises where the device is placed by generating instructions for regulation. Temperature controllers represent the sink element where the information is gathered and locally analysed. Two different types of information are considered: static and dynamic information. Static data is related to one-time information, such as location of the nodes or type of the nodes. Dynamic data is related to time variable parameters such as temperature and energy cost.

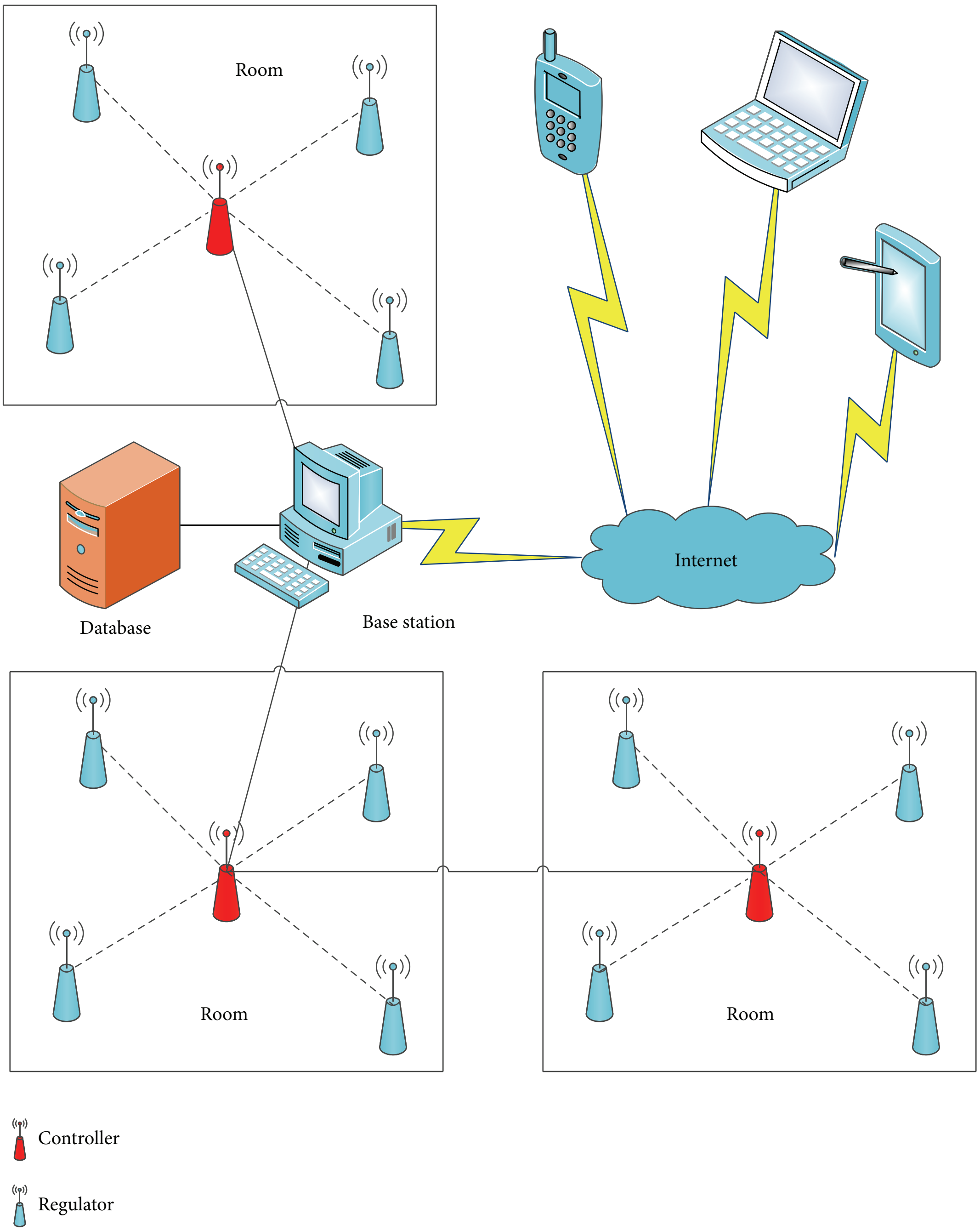

Figure 1: System architecture.

Deployment diagram of the FIRT framework is given in Figure 2. As illustrated, the temperature controller consists of the following elements: local database, rules (which can be based on ontologies [12]), and sensors. Client should be able to remotely configure the rules so the controller can meet the temperature goals. The rules can include, for example, temperature levels for several time intervals during different days. The local database at the controller can be used in order to store information regarding the temperature readings, as well as the static information such as node type and node location. The database can further contain information on the size of the room being monitored, the number of expected visitors, and other factors that can influence the temperature changes. All information that cannot be obtained from the nodes, such as electricity cost, can be obtained from the server. Furthermore, the server contains a database that can be used by the controller to store historical data.

*2.1. Temperature Optimization Framework.* Finally a temperature controller needs to make decisions by generating a set of instructions for the different nodes. Hence, a regulation method is required by the controllers in order to achieve distributed decision regarding the temperature difference that needs to be achieved in a certain point in order to satisfy the required overall temperature.

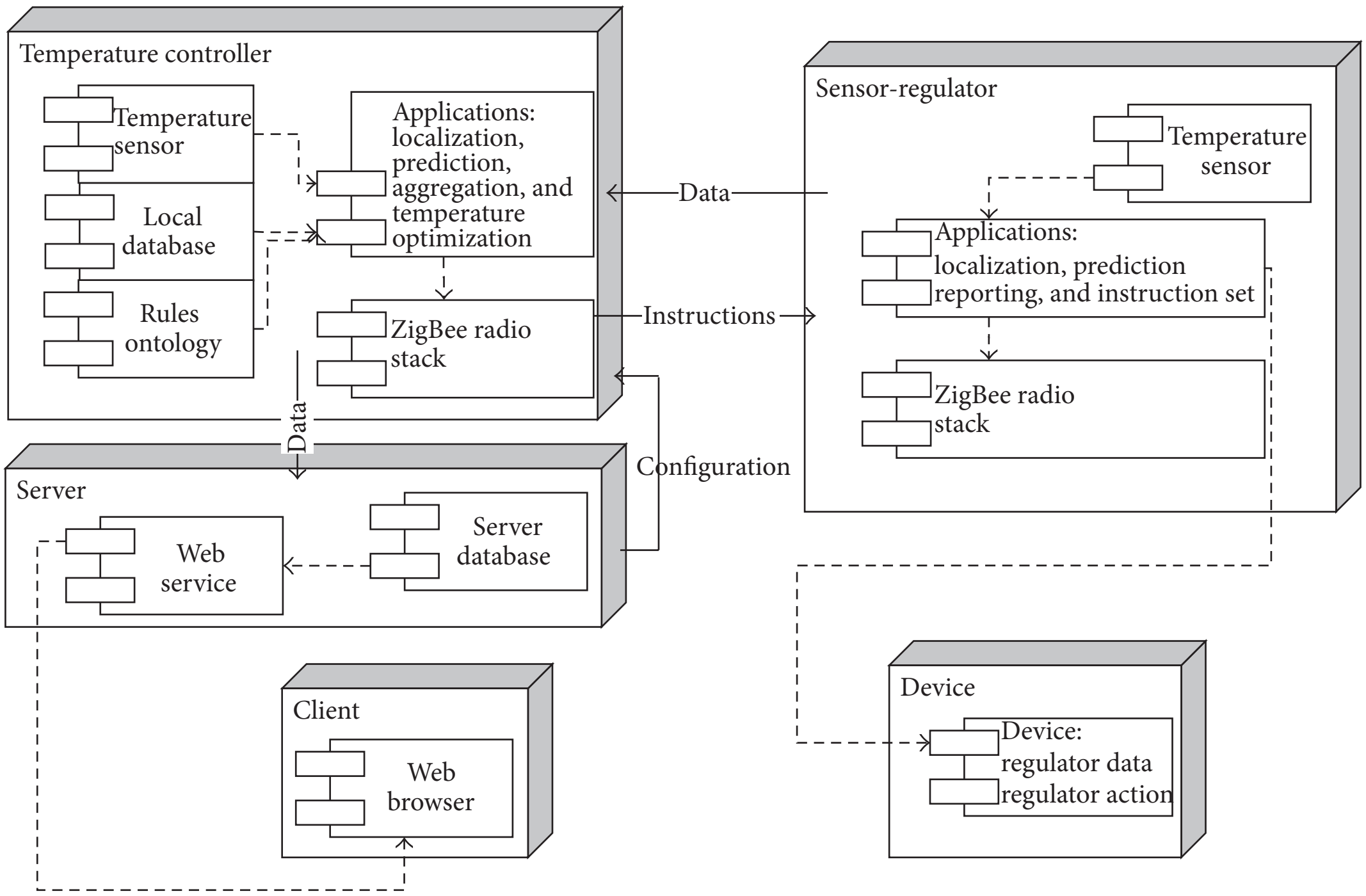

Figure 2: Deployment diagram of the system.

In order to provide a self-organized and cost-effective solution, the WSN must provide the following functionalities: self-localization, nodes clustering, data prediction, distributed decision making, quality of service (QoS), and so forth [13, 14]. Sensor readings are useless if the location where they are measured is not known or if the location is wrong. Thus, a suitable localization algorithm should be implemented in order to discover the location of the nodes. This is important because manual recording of the nodes positions is a very time-consuming solution and prone to errors. The location information will be used by the controller device to deduct the temperature set point that each of the heater shall be commanded to, so that the temperature dissipation from all heaters gives the wanted temperature at the controller's location and the controlled premises overall (elaborated in Section 4). After nodes discovering phase, nodes can be divided into clusters. There are many algorithms purposed in the literature for optimal nodes clustering. In our scenario, we assume that nodes deployed in close proximity to each other belong to the same cluster.

After these two phases (discovering and forming the clusters), we consider that a WSN is established. Nodes can start measuring the temperature and forward the measured readings to the final destination (sink node). In order to save energy, algorithms for data prediction should be implemented on both sides: node and sink (elaborated in Section 5).

The temperature measurements are analysed by generation of temperature gradients. The temperature gradient indicates the direction and the rate at which the temperature changes within a particular location. The dynamic information such as temperature gradients, time of the day, expected visitors, and electricity cost and the static information such as node type and node location represent an input to the temperature optimization method.

The location of the wireless nodes together with the temperature prediction methods allows the calculation of the temperature gradient. Several methods can be used in order to rank or assign weights to the different types of input such as multiple attribute decision-making algorithms, genetic algorithms, analytic hierarchy process, and fuzzy logic. Fuzzy logic is suitable as fuzzy judgement matrices used for the comparative analysis are close to the way the humans reflect and are very easy to implement. The fuzzy inference system can be based on the standard Mamdani or Sugeno methods. For each parameter that needs to be taken into consideration membership function needs to be defined. Moreover, the rules that need to be applied need to be carefully chosen. The actual definition of the membership functions and the rules will be considered in our future work.

After the collected data is analysed, the controller sends instructions to the nodes (regulators). In order to determine the appropriate actions the system should perform (open window, increase radiator temperature, etc.), a mathematical model is needed to calculate the heat transfer rate in a certain object. Choosing the most suitable model depends on the building where the WSN is deployed; that is, old buildings and buildings constructed following the latest recommendations and directives would have different models due to different type of isolation. The following parameters should be considered: heat resistance which depends on the thickness of the material, heat transmission coefficient which depends on the type of the material, and the temperature difference on the opposite sides of the walls, as well as the area of the walls.

## 3. ZigBee Overview and WSN-FIRT Network Topology

ZigBee is a low-cost, low power, wireless mesh network standard [2]. The low cost allows the technology to be widely deployed in wireless control and monitoring applications (in fields such as home automation, health care, and temperature control). Low power usage allows longer life with smaller batteries. Mesh networking provides high reliability and more extensive range. ZigBee operates in the industrial, scientific, and medical (ISM) radio bands (e.g. 868 MHz in Europe, 915 MHz in the USA and Australia, 2.4 GHz in most jurisdiction sworldwide). Data transmission rates vary from 20 kbps in the 868 MHz frequency band to 250 kbps in the 2.4 GHz frequency band.

ZigBee builds upon the physical layer and medium access control defined in the IEEE standard 802.15.4 (2003 version) for low-rate wireless personal area networks (WPANs). The specification completes the standard by adding four main components: network layer, application layer, ZigBee Device Objects (ZDOs), and manufacturer-defined application objects which allow for customization and favor total integration.

Besides adding two high-level network layers to the underlying structure, the most significant improvement is the introduction of ZDOs. These are responsible for a number of tasks, which include keeping of device roles, management of requests to join a network, and device discovery and security.

Due to the fact that ZigBee nodes can go from sleep to active mode in 30 ms or less, the latency is low and devices can be responsive, particularly compared to Bluetooth wake-up delays, which are typically around three seconds. Since ZigBee nodes can sleep most of the time, average power consumption can be low, resulting in long battery life.

The ZigBee network layer natively supports both star and tree network topologies and generic mesh networks. Every network must have one coordinator device, tasked with its creation of the control of its parameters and basic maintenance. Within star networks, the coordinator must be the central node. Both trees and meshes allow the use of ZigBee routers to extend communication at the network level. The routers need to be constantly active, listening for network traffic. Therefore, it is normally assumed that the routers are mains powered sensors/devices. The battery devices are assumed to be sleeping and only waking up and polling for data periodically or on demand (upon user interaction).

The network topology should adapt to the characteristics of the controlled premises. For the proposed WSN-FITR we consider two different topologies: star- and cluster-based topologies. An example of a star topology is illustrated in Figure 3 where one classroom is illustrated. In this case there is a single temperature controller that is responsible for controlling the temperature at the different nodes. The star topology is most appropriate for small areas where there are no major obstacles so that the signal from the nodes does not fade in high extent. For large premises, cluster-based topology is preferred over mesh as in the latter higher energy is required at the nodes due to the fact that each node transmits its own readings and the readings of other nodes. Figure 4 illustrates the proposed WSN-FITR for one floor in a commercial center where there are several temperature clusters that are controlling a set of nodes. The decision regarding the temperature regulation is distributed among all controllers. There is one main controller that has wired connection to the server in order to retrieve/store information towards the database at the server.

## 4. Localization and Clustering in Indoor WSN

Many algorithms have been proposed for ZigBee-based WSN localization. Most of them consider a WSN deployed in outdoor environment where GPS signals are available and a global map of the network can be easily achieved using well-known techniques for localization [15, 16]. On the other side, indoor localization methods should consider different characteristics of the indoor surroundings where WSN is installed. Finding position of indoor WSN is more challenging since GPS signal is heavily attenuated by building structures such as walls and roofs and there is absence of line of sight to some satellites [17]. With only few exceptions, the distances between the nodes of the network are necessary to be known for accurate location prediction. Different techniques are used to obtain the distances:

(i) RSSI (received signal strength indicator),

(ii) ToA (time of arrival),

(iii) AoA (angle of arrival),

(iv) TDoA (time difference of arrival).

The techniques based on RSSI are easier to implement and do not require additional hardware, as all standard wireless devices possess features for measuring this value. But finding the relationship between the signal strength attenuation and the transmission distance in indoor environments is not a trivial task [18, 19]. In the indoor environment, the moving objects inside the building can cause reflection, diffraction, or absorption of the radio signals. Thus the algorithms are more prone to errors due to multipath phenomenon. Additionally, many other characteristics of indoor environments have to be considered like temperature and humidity variations, orientation of antenna, furniture rearrangements, presence of human beings, and so forth.

Indoor localization methods can be divided into two main categories [20]:

(i) deductive (mathematical) methods, that take into account only the physical properties of signal propagation. They require the positions of the access points, radio propagation model and map of the environment,

(ii) inductive (fingerprinting) methods, that require a previous training phase where the system learns the

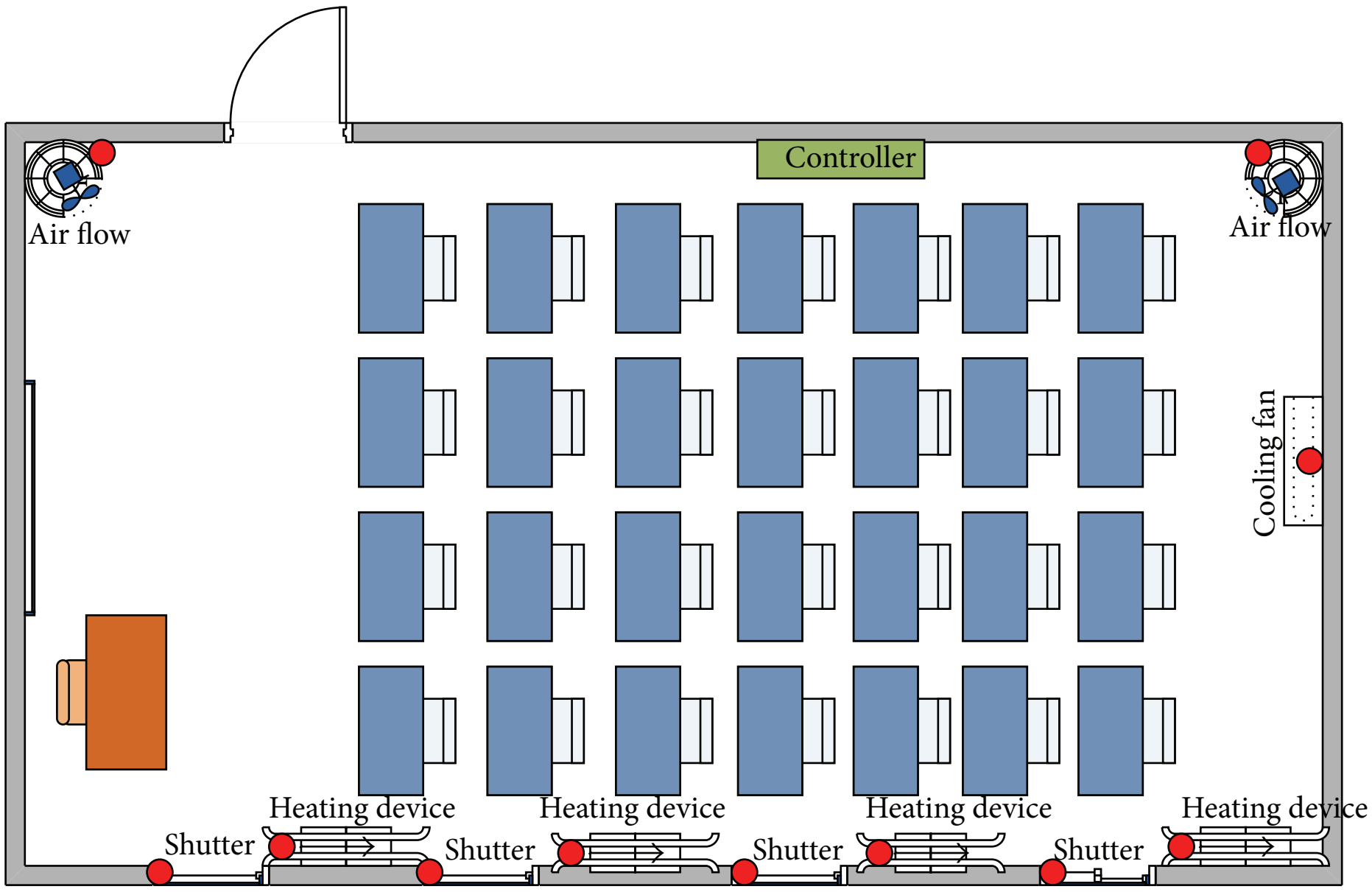

FIGURE 3: Star architecture for a typical classroom.

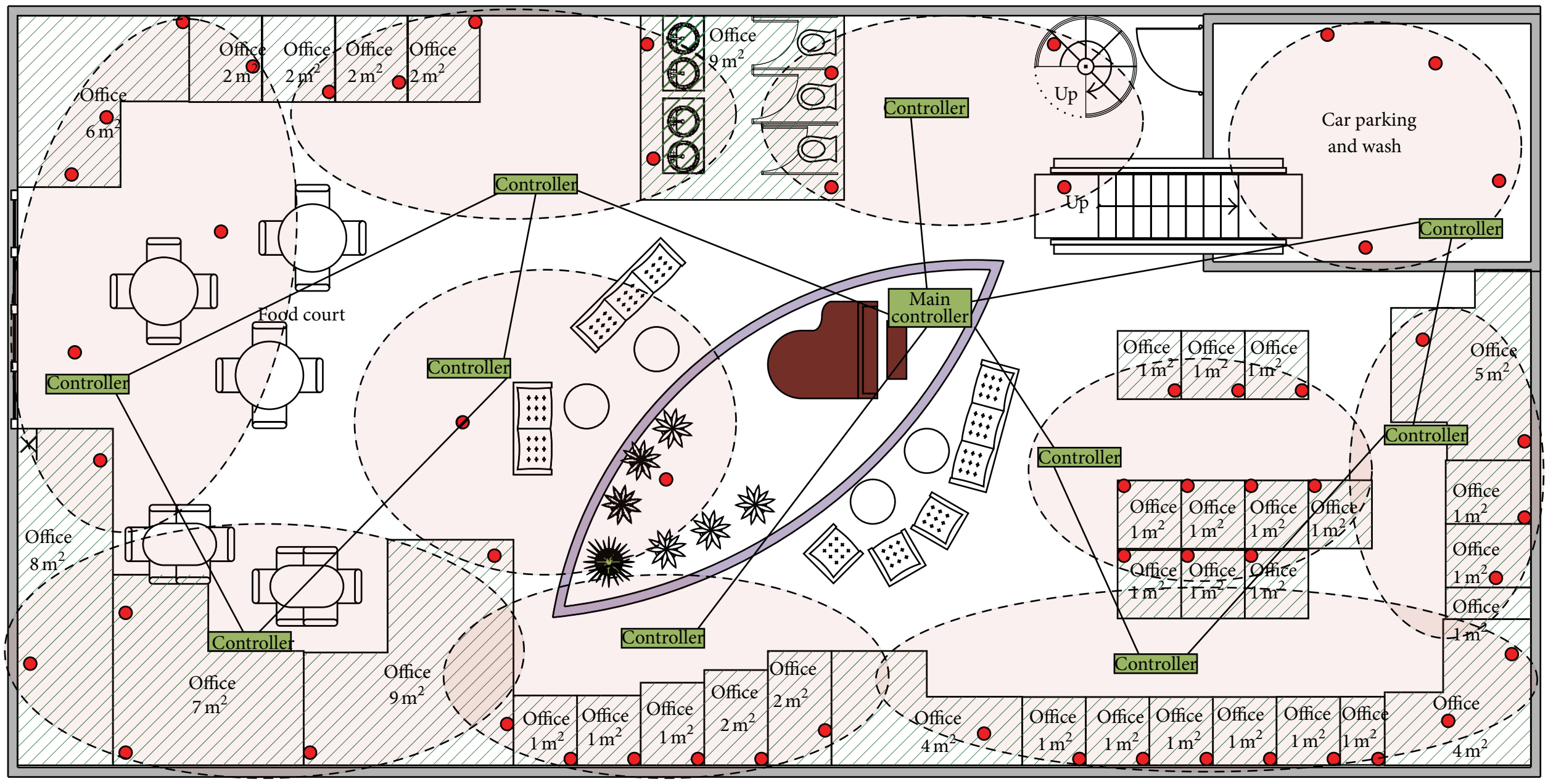

FIGURE 4: Clustered architecture for one floor in a shopping center.

RSS in each location. This phase can be very time-consuming. In the next (positioning) phase, different matching algorithm can be used in order to find the unknown location.

In [20] the authors present an algorithm that combines the advantages of both deductive and inductive methods. This hybrid method reduces the training phase without a loss of precision. In [21] several matching fingerprinting algorithms are investigated: the nearest neighbor (NN) algorithm, the K-weighted nearest neighbor (KWNN) algorithm, and the probabilistic approach based on the kernel method. Through simulations it has been shown that KWNN algorithm has the best indoor positioning result.

Since localization is very crucial in our WSN-FITR, the algorithm should be selected very carefully in accordance with the characteristic of the environment where the network should be deployed. If there are many walls and obstacles in the environment, the deductive methods should be avoided because they estimate the position mathematically. When there are multiple access points and few walls in the environment, inductive methods are not necessary as the training phase can be very expensive.

With the expansion of pervasive computing, most of the service-based applications are expected to be contextual awareness and location dependent. Still, indoor localization is pioneering and finding appropriate methods will be a challenge in the next years.

After determining the location, sensor nodes in WSN can be geographically grouped into clusters. In each cluster one representative node (cluster head) is chosen to coordinate member nodes. The main advantages of WSN clustering are not only to prolong the WSN lifetime but also to establish collaboration between cluster members in order to provide data aggregation and more accurate reports about the region they sense. Many algorithms have been proposed in the literature for WSN clustering [22, 23].

## 5. Reductions of Data Transmissions

By reporting data measurement at each interval, WSN nodes consume a great deal of energy, which reduces network lifetime and creates sufficient communication overhead. Several techniques have been developed to overcome these problems, that is, to lower the communication overhead and to increase energy saving. Most of them consider reducing the number of radio transmissions.

Three main paradigms can achieve reduction of radio transmissions:

(i) *data compression,* where well-known compression techniques are used to compress consecutive measurements. This approach is useful only if the WSN application does not require the data in real time and can be achieved regardless of the network topology,

(ii) *in-network processing,* where data are processed on their way to the sink. This method is usually performed when summarization functions or other queries are needed. It is appropriate only for mesh-based, cluster-based, or hybrid-based network topologies but cannot be implemented for star-based network topology,

(iii) *data prediction,* where different prediction methods are used for predicting the next sensor readings. Here, each node runs a filter (or a model) that estimates the next sensor reading. The sink runs exactly the same models for each sensor in the network and makes the same predictions. This approach is known as dual prediction scheme (DPS).

For the WSN-FITR system we need the sensor measurements up-to-date; hence data compression is not an appropriate solution. Regarding network topology, we can choose among data prediction and innetwork processing. If the network topology of WSN-FITR is star based, we should apply data prediction methods. For different topologies we should consider innetwork processing or combination of both.

In order to compare these techniques for data reduction, different algorithms were implemented in MATLAB. For the evaluation, a set of experimental data from Intel Lab [24] was used. The 54 Mica2Dot sensors deployed in the laboratory were equipped with weather boards and measured temperature once every 31 seconds. The measurements were collected between February 28 and April 5, 2004. We run the simulations for 50 different error margins $E_{max}$ (ranging from 0.1°C to 5°C).

*5.1. Data Prediction.* The most appropriate models (filters) for DPS are based on time-series forecasting: moving average (MA), autoregressive (AR) model [25], autoregressive moving average (ARMA) [26], least mean square (LMS) [27], and LMS with variable step size (LMS-VSS) [28]. We implemented and evaluated LMS, LMS-VSS, second-, fourth-, and tenth-order MA and ARMA. Figure 5 shows the reduction gain for each of these algorithms simulated on two nodes form the Intel Berkeley Research Lab network [24]. The metric is the reduction of transmissions in percentage (Figure 5, upper) and the difference between the predicted and the true value (Figure 5, lower), that is, mean square error (MSE). The horizontal axis represents the value of the threshold of $E_{max}$.

In Figures 6 and 7, the results of the algorithms are represented by different colors: blue for MA(2), yellow for MA(4), red for MA(10), and green for ARMA. The results show that, concerning the reduction of transmissions, the ARMA algorithm is constantly better than the other two (Figure 6). MA(2) is second in this regard, and MA(4) is slightly behind. In most of the results from the simulations we have done, the difference is greater for threshold values in the [0, 1.5] range. However, there is very little difference for values greater than 1.5.

Another conclusion, concerning the performance of the MA algorithm that can be drawn from the results, is that both MSE and percentage of sent data are positively correlated with the order of the MA algorithm. This is more clearly visible on Figure 7, where only the three MA algorithms (MA(2), MA(4), and MA (10)) are compared. Although ARMA always performs better than the other techniques, this algorithm uses 20 consecutive readings in order to generate the prediction model, while the others (MA, LMS, and LMS-VSS) operate on real time.

*5.2. In-Network Processing in WSN-FITR.* If the WSN is deployed on vast region, the radio signals would be far from the sink node and multihop routing is needed for data to reach its destination.

In order to calculate the reduction in this case, we divided Intel network [24] into clusters (Figure 8). Four algorithms were used for evaluation, MA(2), MA(4), LMS, [27] and LMS-VSS [28]. The clustering parameter was geographic position, that is, the Euclidian distance. We assume that each sensor sends its reading to the cluster head, which is responsible for resending the reading to the sink. As a result, each reading is sent twice, except for the readings taken at the cluster head.

Figure 9 shows the reduction for cluster containing nodes: 1, 33, 34, 35, 36, and 37. For error margin of 0.5°C, LMS-VSS shows an average gain of 2.5% compared to LMS algorithm, MA(4) is better than LMS-VSS by 4%, while MA(2) outperforms MA(4) by approximately 1%. Additional reduction can be achieved if cluster head performs summarization function (Average, Minimum, Maximum, etc.) and forwards only the calculated aggregate to the sink. In this case, the data reduction is far greater (97% reduction of the total messages sent for the given error margin of 0.5°C).

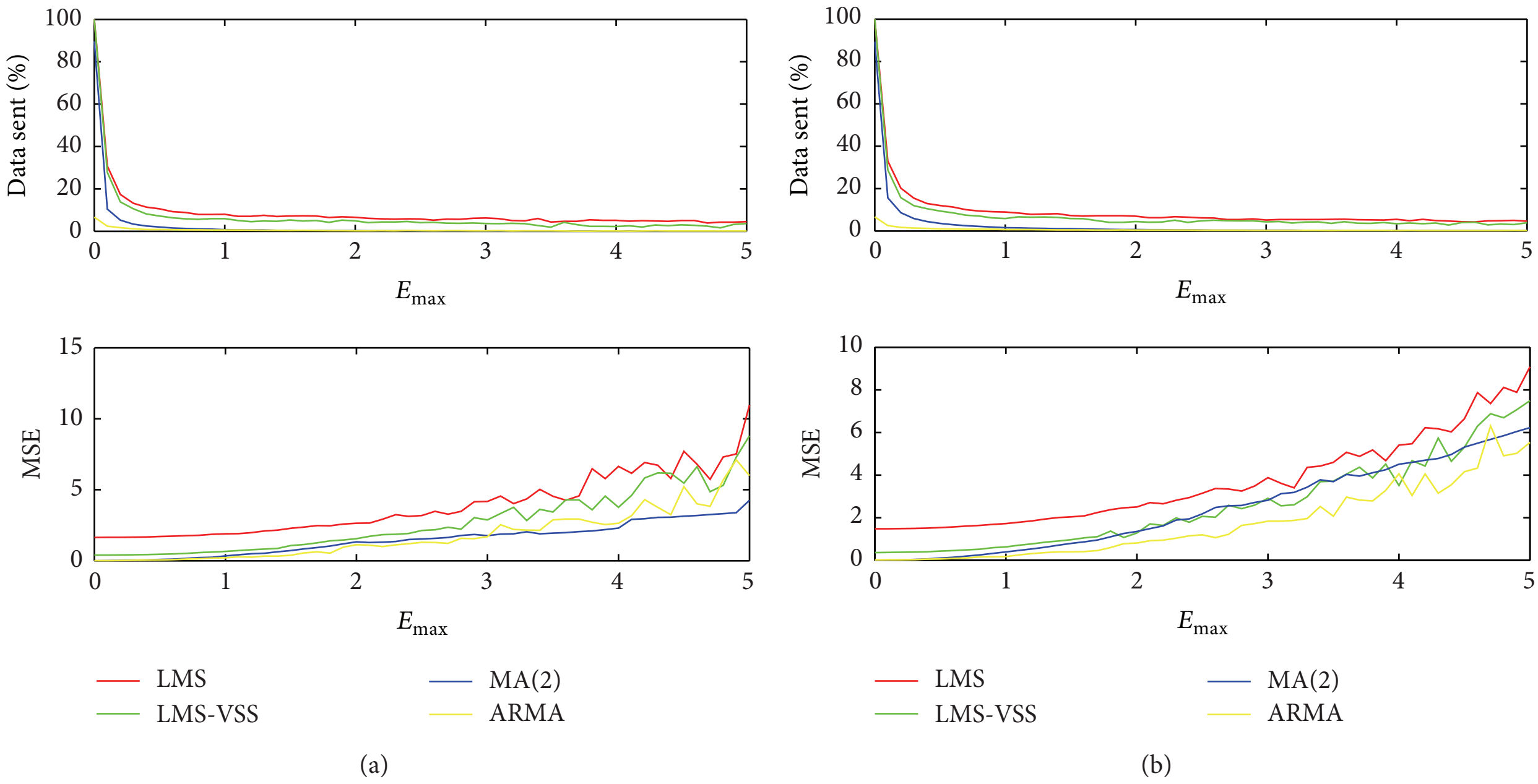

FIGURE 5: Data reduction for different algorithms for node 13 (a) and node 49 (b).

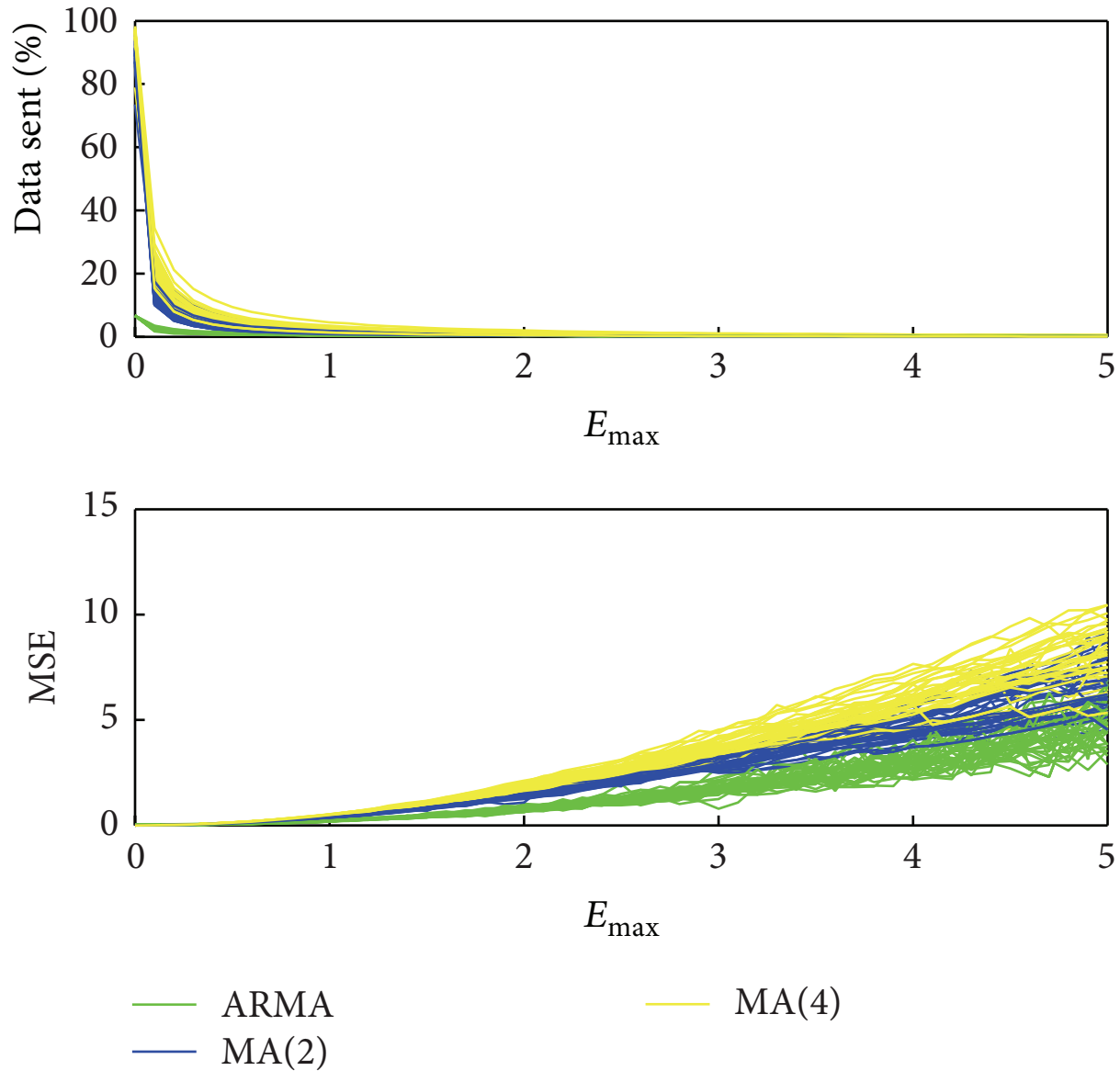

FIGURE 6: Results from simulations performed on 35 different nodes.

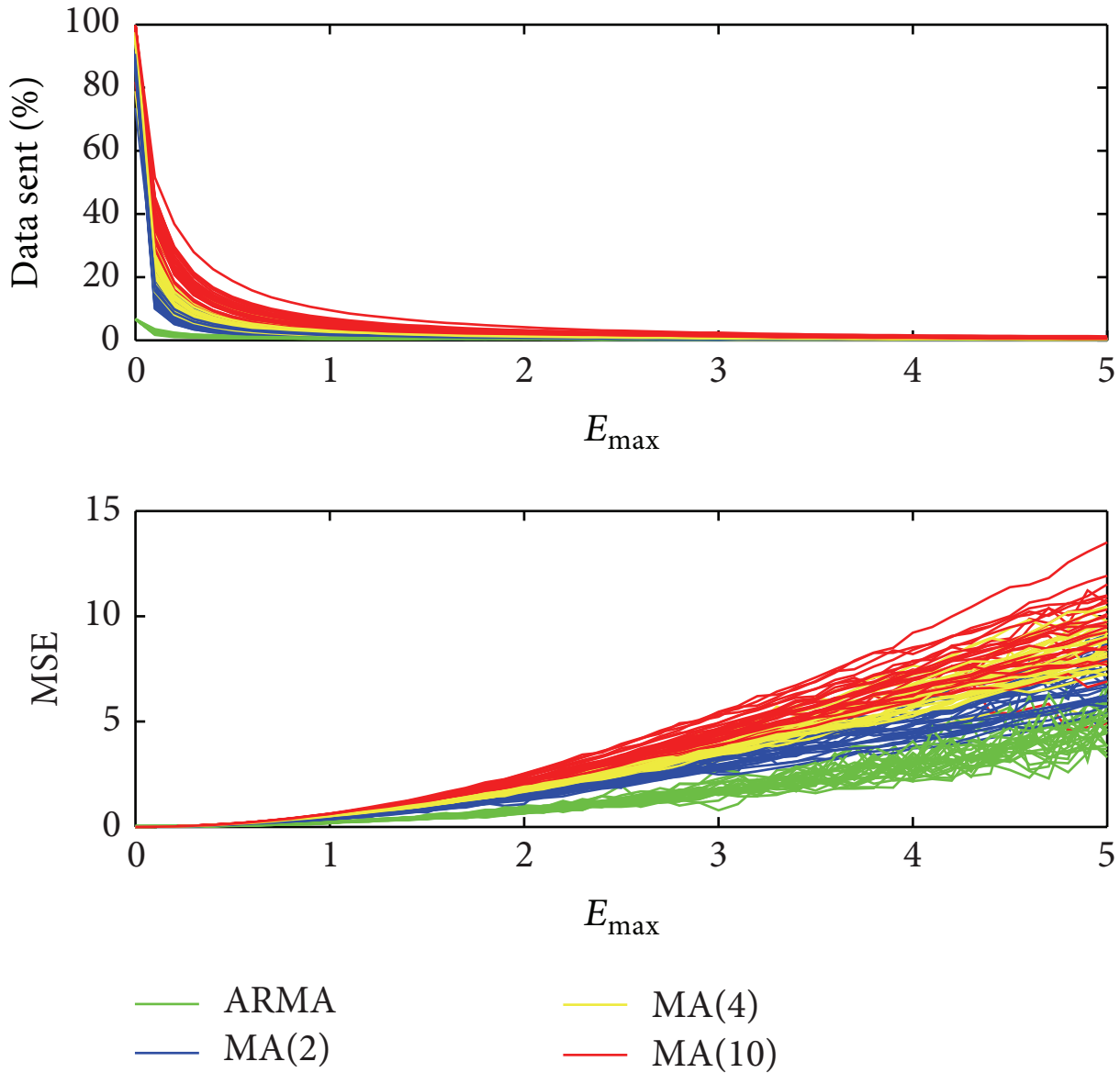

FIGURE 7: Results from simulations performed on 30 different nodes.

From the results, it can be concluded that, for these datasets, the MA(2) always performs better compared with other prediction techniques.

## 6. Conclusion

In this paper we have proposed a WSN framework for indoor temperature regulation. We provided an overview of systems architecture and in detail elaborated designing guidelines for ZigBee-based network. In order to reduce the energy consumption, we proposed a data reduction strategy based on dual prediction scheme that uses different methods for time-series forecasting. Through simulations on real world dataset we showed that these filters can achieve reduction of data transmissions of more than 90%.

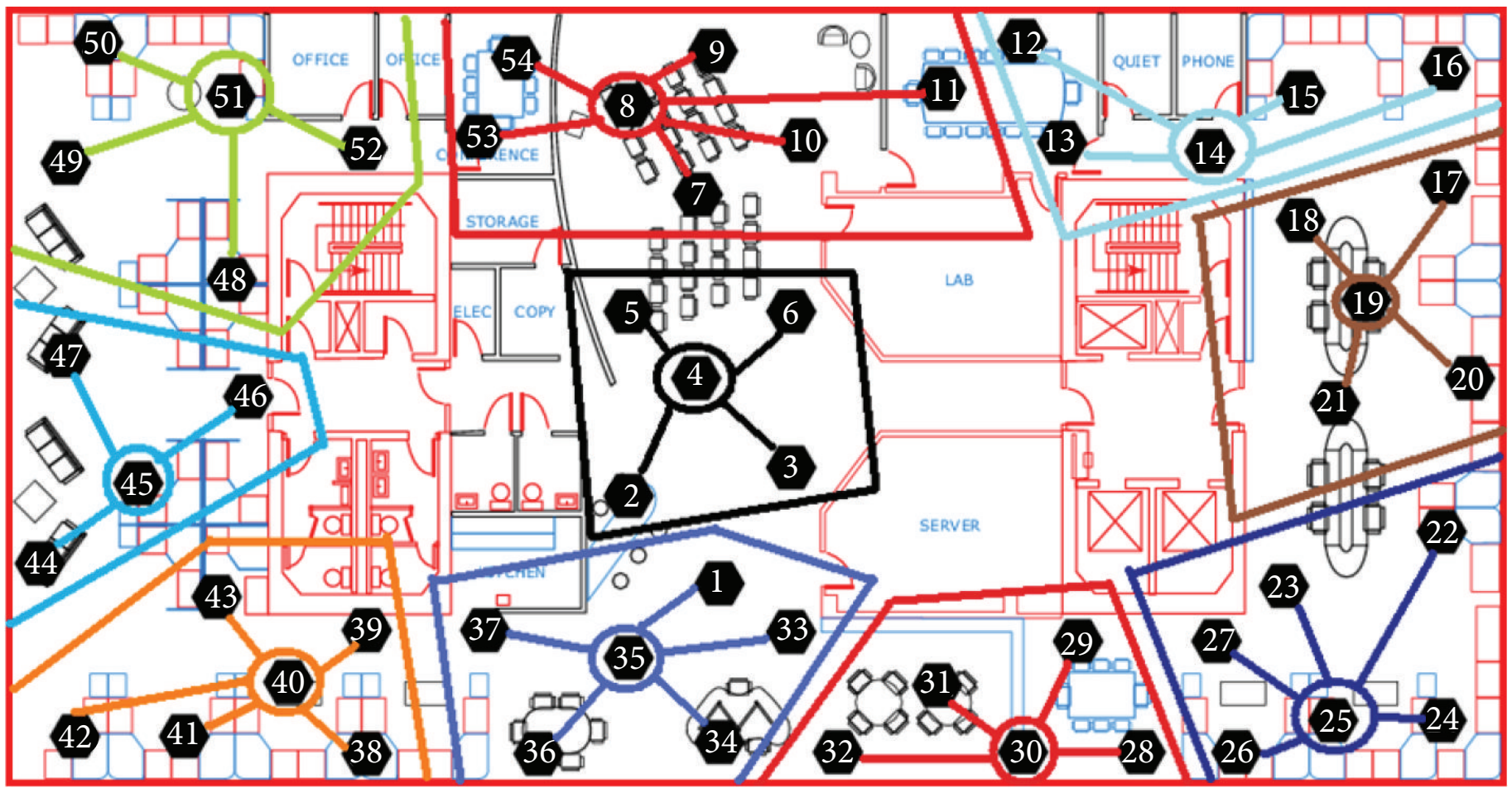

FIGURE 8: A clustered view of the Intel Lab [24].

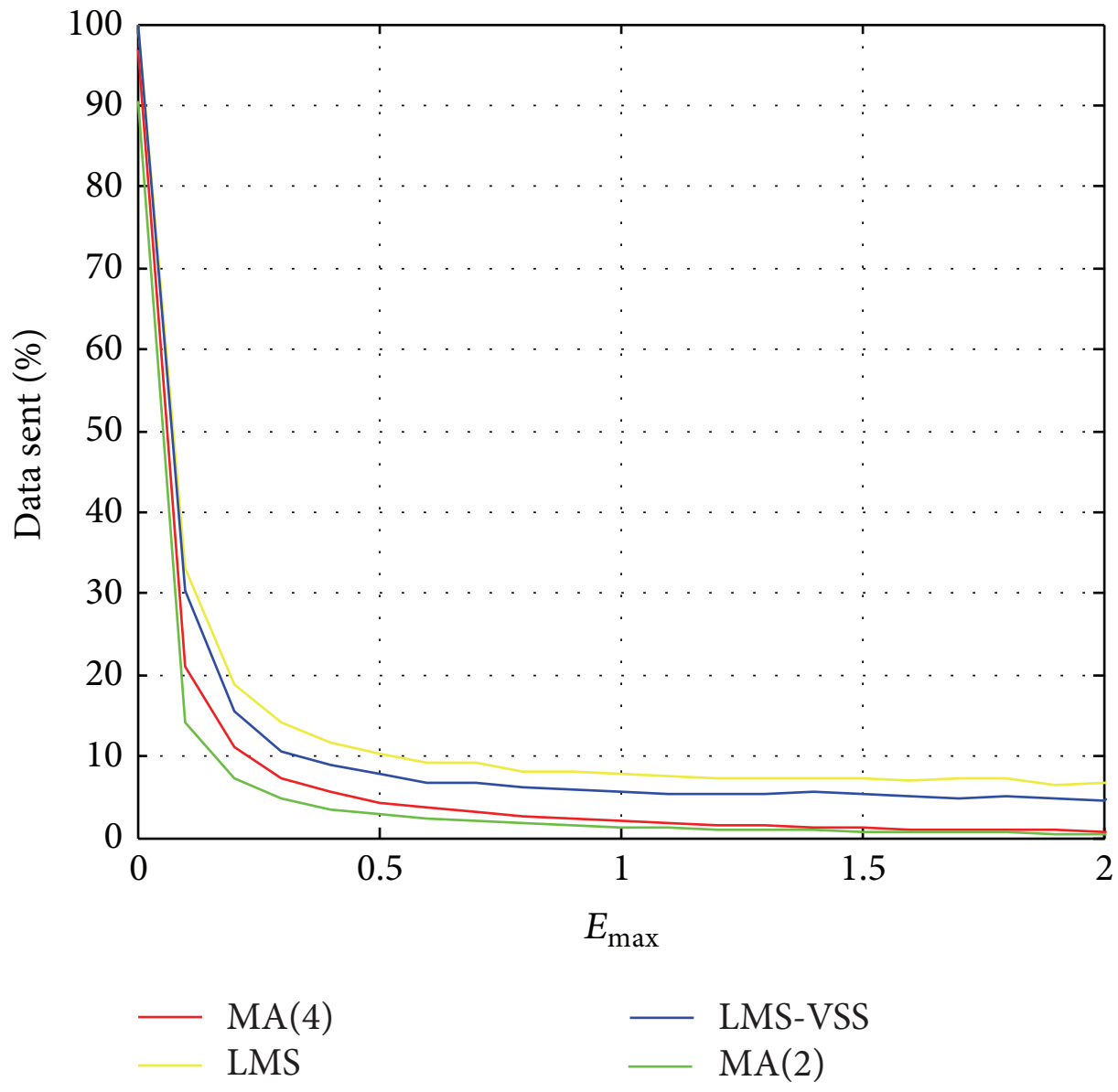

FIGURE 9: Data reduction in the cluster containing the nodes: 1, 33, 34, 35, 36, and 37.

For future work, we intend to investigate in the above-mentioned data prediction techniques using different datasets with different measured parameters, like light intensity or humidity, since they can affect the temperature parameter. We also plan to explore machine learning techniques for choosing the best technique. Furthermore, for distributed decision making, fuzzy logic is very suitable to be implemented in constrained WSNs, since, for temperature regulation, there are only a few important parameters that hold most of the information and make rule pruning possible.

## Conflict of Interests

The authors declare that there is no conflict of interests regarding the publication of this paper.

## References

[1] R. Mittal and M. P. S. Bhatia, "Wireless sensor networks for monitoring the environmental activities," in *Proceedings of the IEEE International Conference on Computational Intelligence and Computing Research (ICCIC '10)*, pp. 1–5, December 2010.

[2] "ZigBee Alliance," http://www.zigbee.org/.

[3] C. Zhang, M. Zhang, Y. Su, and W. Wang, "Smart home design based on ZigBee wireless sensor network," in *Proceedings of the 7th International ICST Conference on Communications and Networking in China (CHINACOM '12)*, pp. 463–466, August 2012.

[4] A. A. Khan and H. T. Mouftah, "Energy optimization and energy management of home via web services in smart grid," in *Proceedings of the IEEE Electrical Power and Energy Conference (EPEC '12)*, pp. 14–19, IEEE, Ontario, Canada, October 2012.

[5] V. Boonsawat, J. Ekchamanonta, K. Bumrungkhet, and S. Kittipiyakul, "Xbee wireless sensor networks for temperature monitoring," in *Proceedings of the 2nd ECTI-Conference on Application Research and Development (ECTI-CARD '10)*, Pattaya, Chonburi, Thailand, May 2010.

[6] "Arduino Platforms," http://www.arduino.cc/.

[7] "Xbee Wireless RF Modules," http://www.digi.com/xbee/.

[8] C. C. Castello, R. X. Chen, J. Fan, and A. Davari, "Context aware wireless sensor networks for smart home monitoring," *The International Journal of Autonomous and Adaptive Communications Systems*, vol. 6, pp. 99–114, 2013.

[9] M. Demirbas, K. Y. Chow, and C. S. Wan, "INSIGHT: internet-sensor integration for habitat monitoring," in *Proceedings of the Advanced Experimental Activities on Wireless Networks and Systems (EXPONWIRELESS) Workshop (as part of WOWMOM 2006)*, pp. 553–558, Buffalo, NY, USA, June 2006.

[10] B. Stojkoska and D. Davcev, "Web interface for habitat monitoring using wireless sensor network," in *Proceedings of the 5th International Conference on Wireless and Mobile Communications (ICWMC '09)*, pp. 157–162, IEEE, August 2009.

[11] A. Caracas, F. Mller, O. Sundstrm, C. Binding, and B. Jansen, "VillaSmart: wireless sensors for system identification in domestic buildings," in *Proceedings of the IEEE Innovative Smart Grid Technologies Europe (ISGT Europe '13)*, Copenhagen, Denmark, October 2013.

[12] A. Rossello-Busquet, J. Soler, and L. Dittmann, "A novel home energy management system architecture," in *Proceedings of the UKSim 13th International Conference on Modelling and Simulation (UKSim '11)*, pp. 387–392, April 2011.

[13] B. Yahya and J. Ben-Othman, "Energy efficient and QoS aware medium access control for wireless sensor networks," *Concurrency Computation Practice and Experience*, vol. 22, no. 10, pp. 1252–1266, 2010.

[14] J. Ben-Othman, K. Bessaoud, A. Bui, and L. Pilard, "Self-stabilizing algorithm for efficient topology control in Wireless Sensor Networks," *Journal of Computational Science*, vol. 4, no. 4, pp. 199–208, 2013.

[15] B. Stojkoska, I. Ivanoska, and D. Davcev, "Wireless sensor networks localization methods: multidimensional scaling versus semidefinite programming approach," in *ICT Innovations 2009*, pp. 145–155, Springer Berlin, Heidelberg, Germany, 2010.

[16] B. R. Stojkoska and D. Davcev, "MDS-based Algorithm for Nodes Localization in 3D Surface Sensor Networks," in *Proceedings of the 7th International Conference on Sensor Technologies and Applications (SENSORCOMM '13)*, pp. 44–50, 2013.

[17] L. Cheng, C. Wu, Y. Zhang, H. Wu, M. Li, and C. Maple, "A survey of localization in wireless sensor network," *International Journal of Distributed Sensor Networks*, vol. 2012, Article ID 962523, 12 pages, 2012.

[18] A. Awad, T. Frunzke, and F. Dressler, "Adaptive distance estimation and localization in WSN using RSSI measures," in *Proceedings of the 10th Euromicro Conference on Digital System Design Architectures, Methods and Tools (DSD '07)*, pp. 471–478, August 2007.

[19] A. Faheem, R. Virrankoski, and M. Elmusrati, "Improving RSSI based distance estimation for 802.15.4 wireless sensor networks," in *Proceedings of the IEEE International Conference on Wireless Information Technology and Systems (ICWITS '10)*, August 2010.

[20] J. Lloret, J. Tomas, M. Garcia, and A. Canovas, "A hybrid stochastic approach for self-location of wireless sensors in indoor environments," *Sensors*, vol. 9, no. 5, pp. 3695–3712, 2009.

[21] S.-S. Jan, L.-T. Hsu, and W.-M. Tsai, "Development of an indoor location based service test bed and geographic information system with a wireless sensor network," *Sensors*, vol. 10, no. 4, pp. 2957–2974, 2010.

[22] S. Zhang, J. Fan, J. Jia, and J. Wang, "An efficient clustering algorithm in wireless sensor networks using cooperative communication," *International Journal of Distributed Sensor Networks*, vol. 2012, Article ID 274576, 11 pages, 2012.

[23] A. A. Abbasi and M. Younis, "A survey on clustering algorithms for wireless sensor networks," *Computer Communications*, vol. 30, no. 14-15, pp. 2826–2841, 2007.

[24] "Intel lab data.Web Page," http://db.lcs.mit.edu/labdata/labdata.html.

[25] Y.-A. le Borgne, S. Santini, and G. Bontempi, "Adaptive model selection for time series prediction in wireless sensor networks," *Signal Processing*, vol. 87, no. 12, pp. 3010–3020, 2007.

[26] J. Lu, F. Valois, M. Dohler, and M.-Y. Wu, "Optimized data aggregation in WSNs using adaptive ARMA," in *Proceedings of the 4th International Conference on Sensor Technologies and Applications (SENSORCOMM '10)*, pp. 115–120, Venice, Italy, July 2010.

[27] S. Santini and K. Romer, "An adaptive strategy for quality-based data reduction in wireless sensor networks," in *Proceedings of the 3rd International Conference on Networked Sensing Systems (INSS '06)*, pp. 29–36, Chicago, IL, USA, June 2006.

[28] B. Stojkoska, D. Solev, and D. Davcev, "Data prediction in WSN using variable step size LMS algorithm," in *Proceedings of the 5th International Conference on Sensor Technologies and Applications*, Nice, France, 2011.